\documentclass[twocolumn,showpacs,amsmath,amssymb,aps,prd,floats]{revtex4-1}
\usepackage{bm}
\usepackage{graphicx}
\usepackage{times}
\usepackage{dcolumn}
\input{colordvi.tex}

\begin{document}

\title{Diffuse PeV neutrinos from  {EeV cosmic ray sources:\\ semi-relativistic} hypernova remnants in star-forming galaxies}

\author{Ruo-Yu Liu$^{1,2,5,*}$}
\author{Xiang-Yu Wang$^{1,5}$}
\author{Susumu Inoue$^{2}$}
\author{Roland Crocker$^{3}$}
\author{Felix Aharonian$^{4,2}$}

\affiliation{
$^1$School of Astronomy and Space Science, Nanjing University, Nanjing, 210093, China\\
  $^2$Max-Planck-Institut f\"ur Kernphysik, 69117 Heidelberg, Germany\\
  $^3$Research School of Astronomy \& Astrophysics, Australian National University, Weston Creek, ACT 2611, Australia\\
  $^4$Dublin Institute for Advanced Studies, 31 Fitzwilliam Place, Dublin 2, Ireland\\
  $^5$Key Laboratory of Modern Astronomy and Astrophysics (Nanjing University), Ministry of Education, Nanjing 210093, China\\
  $^*$Fellow of the International Max Planck Research School for Astronomy and Cosmic Physics at the University of Heidelberg (IMPRS-HD)
}

\begin{abstract}
We argue that the excess of sub-PeV/PeV neutrinos recently reported by IceCube
could plausibly originate through pion-production processes in the same sources 
responsible for cosmic rays (CRs) with energy above the second knee around $10^{18}\,$eV.  
 {The pion production efficiency for escaping CRs that produce PeV neutrinos is required to be $\gtrsim 0.1$ in such sources.}
On the basis of current data, we identify semi-relativistic hypernova remants as possible sources that satisfy the requirements.
By virtue of their fast ejecta, such objects can accelerate protons to EeV energies,
which in turn can interact with the dense surrounding medium during propagation in their host galaxies
to produce sufficient high-energy neutrinos via proton--proton ($pp$) collisions.
Their accompanying gamma ray flux  {can remain} below the diffuse isotropic gamma ray background
observed by the {\it Fermi} Large Area Telescope (LAT).
 {
In order to test this scenario and discriminate from alternatives, the density of target protons/nuclei
and the residence time of CRs in the interacting region are crucial uncertainties that need to be clarified.
As long as the neutrinos and EeV CRs originate from the same source class,
detection of $\gtrsim 10\,$PeV neutrinos may be expected within 5-10 years' operation of IceCube.
Together with further observations in the PeV range, the neutrinos can help in revealing the currently unknown sources of EeV CRs.
}
\end{abstract}


\maketitle 
\section{Introduction}
 {Observations of high-energy neutrinos 
have important implications for understanding the
origin of PeV-EeV cosmic rays (CRs),}
because the collisions of hadronic CRs with background nuclei 
or photons 
produce, among other particles, charged mesons whose decay products include neutrinos: 
($\pi^{+}\rightarrow e^{+}\nu_\mu \bar{\nu}_\mu \nu_e$,
$\pi^{-}\rightarrow e^{-}\nu_\mu \bar{\nu}_\mu \bar{\nu}_e$). 
Two PeV neutrinos were detected by the IceCube neutrino detector
during the combined IC-79/IC-86 data period \cite{IC13}. 
More recently, follow-up analysis by the IceCube Collaboration
uncovered 26 additional sub--PeV neutrinos \cite{ICSci13}.
%
 {They show that these 28 events in total,}
ranging from 60\,TeV--2PeV, correspond to a 4.3 $\sigma$ excess over  {reasonable expectations for the}
background of $10.6^{+4.5}_{-3.9}$ from atmospheric neutrinos and muons,
corresponding to a single-flavor neutrino flux
of $(1.2\pm 0.4)\times 10^{-8}\, \rm GeV\, cm^{-2}s^{-1}sr^{-1}$ at PeV.

Non-detection of higher energy events implies a
cutoff or a break above 2\,PeV for a hard spectrum with power-law index of $s_\nu=2$. 
Alternatively, it is also compatible with a slightly
softer but unbroken power-law
spectrum with index $s_\nu \simeq 2.2-2.3$ \cite{ICSci13, Anchordoqui13, Winter13}. 

The sky distribution of the 28 events is consistent with
isotropy \cite{ICSci13}, implying an extragalactic origin,
although a fraction of them could come from Galactic sources \cite{Fox13}. 
Several possible scenarios for the extragalactic origin of these
neutrinos have been discussed, including 
that they are `cosmogenic', arising in
$p\gamma$-collisions
between CRs and cosmic background photons, 
or that they are generated within  {CR} sources,
either in $p\gamma$- or $pp$-collisions between CRs and ambient radiation fields or gas
respectively \cite{Prev, Roulet13, Murase13, Anchordoqui13, Winter13}. 
A cosmogenic origin for the IceCube events is
excluded because the predicted PeV flux is well below the observed one \cite{Roulet13}. 
$p\gamma$- or $pp$-collisions inside sources are
more promising for generating sufficient flux. 
Each daughter neutrino typically takes 3 and 5 percent of the parent proton's energy in these two processes respectively \cite{Kelner06}. 
Thus, to produce a 1\,PeV
neutrino, we require a source  located at redshift $z$ to
accelerate protons  {to} $\gtrsim (40-60)\frac{1+z}{2}\,$PeV.  
This is only an order of magnitude lower than the energy of the ``second knee" (4-8$\times 10^{17}\,$eV),
 {where the CRs spectral index steepens from -3.1 to -3.3.
About one to two orders of magnitude higher,
the spectral index flattens from -3.3 to -2.7 at the ``ankle" ($\lesssim 10^{19}$eV). 
Either of these two spectral features may correspond to the transition energy above which
extragalactic CRs dominate over Galactic CRs \cite{Berezinsky06, Katz09}.}
This motivates us to discuss a possible link between sources of these neutrinos and the sources of 
 {CRs with energies above the second knee, hereafter simply ultrahigh energy CRs (UHECRs) in this paper}.
Note that certain kinds of
 {extragalactic accelerators of protons up to $\sim$100\,PeV may be}
sufficient to explain the current observations.
However,
 {our interest here is}
whether a link could exist between the newly-detected neutrinos and UHECRs,
since then these neutrinos could shed some light on the
 {still mysterious}
sources of UHECRs.
We note that the reported flux is quite close to the
 {so-called}
Waxman-Bahcall bound \citep{Waxman98},
 {a benchmark value for the extragalactic neutrino flux based on the UHECR flux,
subject to some assumptions \cite{Rachen00}.}
Alternative constraints on the extragalactic neutrino flux comes from observations of the isotropic background
of multi-GeV gamma-rays, which is at the level of $10^{-7}\rm \, GeV cm^{-2}s^{-1}sr^{-1}$. It provides a robust upper limit since the gamma-rays that are unavoidably co-produced must not overwhelm this flux.
 {If the PeV neutrinos and UHECRs indeed originate from the same sources,}
the neutrino spectrum should extend to $\gtrsim 10\,$PeV without any abrupt cutoff.
This would not conflict with the current IceCube
observations if the neutrino spectrum is softer than $E^{-2.2}$. 
Note
that the source proton spectrum may not necessarily be soft as the
neutrino spectrum, since in some specific scenarios, higher energy
protons can have lower production efficiencies of secondary pions,
and for $p\gamma$ processes,
the neutrino spectrum also depends on the ambient photon spectrum.
Given the
 {likely}
pion-production origin of the reported neutrinos,
an approximate value for the required flux of parent protons $\Phi_p$
 {that escape the source} can be given by
$\varepsilon_\nu^2\Phi_\nu=\frac{1}{6}f_{\pi}(\varepsilon_p^2 \Phi_p)$ \cite{Waxman98, Murase13, factor},
where $\varepsilon_\nu$ and $\varepsilon_p$ are the energies of
the neutrino and proton respectively, and $f_{\pi}$ is
 {the pion-production efficiency via $pp$- or $p\gamma$-collisions of the escaping CRs.}
Thus, sources of UHECRs that also account for the sub--PeV/PeV
neutrinos need to provide a proton flux of $\varepsilon_p^2 \Phi_p
= 6(\varepsilon_\nu^2\Phi_\nu)f_{\pi}^{-1}\simeq 7\times
10^{-8}\,f_{\pi}^{-1} \rm GeV\,cm^{-2}s^{-1}sr^{-1}$ in the
10--100\,PeV energy range. 
This flux corresponds to a local proton energy production rate of
\begin{equation}\label{energybudget}
\dot{W}_{p,0}\simeq \left(\frac{c\xi_z}{4\pi H_0}\right)^{-1}\alpha(\varepsilon_p^2\Phi_p) \simeq 10^{44.5}f_{\pi}^{-1}\, \rm erg\,Mpc^{-3}yr^{-1}
\end{equation}
where $c$ is the speed of light, $H_0$ is the Hubble constant, $\xi_z\simeq 3$ is a factor that accounts for the
contribution from high-redshift sources \cite{Waxman98}, and $\alpha\sim 10-100$ is a factor coming
from normalization of the proton spectrum (e.g., for power-law index of {$s_p=2$}, $\alpha={\rm ln}\,(\varepsilon_{p, \rm
max}/\varepsilon_{p, \rm min})$). 
Note that accelerated protons contribute to the observed CRs only if they can escape from the sources,
while pion-production process at the source would remove energy from accelerated protons.
Thus the 
 {energy production rate of the CRs that escape the source}
can be given by
$\dot{W}_{\rm CR,0}=\dot{W}_{p,0}(1-\xi f_\pi^{\rm PeV})$ with $\xi=f_{\pi}^{UHE}/f_{\pi}^{\rm PeV}$,
 {where $f_{\pi}^{UHE}$ is the pion production efficiency of the escaping UHECRs.}
For comparison, the required local CR energy production rate is
$\sim 10^{45.5}\rm \,
erg\,Mpc^{-3}yr^{-1}$ if the transition from Galactic to extragalactic CRs occurs at the second knee,
and  $\sim 10^{44.5}\rm \,erg\,Mpc^{-3}yr^{-1}$ if the transition occurs at
the ankle for $s_p=2$ \cite{Katz09}.
 {Given the proton energy production rate for a certain class of source,
the pion production effiency needs to be in a certain range
in order to simultaneously account for the observed neutrino flux,
which in turn can constrain the potential sources.}

The rest of this paper is outlined as follows.
First we provide a brief overview of various candidate sources of UHECRs
 {and discuss their potential as PeV neutrino sources}
in Section II.
Then we focus on semi-relativistic hypernovae in star-forming galaxies as a possible source class
that can simultaneously account for the newly discovered sub-PeV/PeV neutrinos and UHECRs
in Section III. In Section IV, we conclude with a discussion of further aspects concerning the proposed scenario.
%

\section{Possible link between PeV neutrinos and UHECRs}
 {
Considering some selected types of sources
that are known to meet the Hillas criterion \cite{Hillas84} for acceleration of UHECRs,
we indicate in Fig.~1 the typical regions that they may occupy
on the plane of $\dot{W}_{p,0}$, the local proton energy production rate,
versus $f_\pi$, the pion-production efficiency of CRs that produce PeV neutrinos.
The black solid line represents the relation between $\dot{W}_{p,0}$ and $f_\pi$ required to reproduce
the observed neutrino flux, with the gray band corresponding to its 1-$\sigma$ confidence interval.
The upper and lower dashed curves represent the local energy production rate $\dot{W}_{\rm CR,0}$
of escaping CRs required to account for the observed UHECRs if the Galactic-extragalactic transition
occurs at the second knee and at the ankle, respectively, for the case $\xi=1$.
The dotted curves are corresponding ones for the case $\xi=0$.
Different values of $\xi$ will result in different sets of the two curves.
Valid sources of UHECRs are expected to be located above the lower curves.
If, in addition, the efficiency of escape of accelerated CRs from the source is high,
they should lie below the upper curves.
}
Note that $\alpha=10$ has been adopted here. A larger $\alpha$ will shift all the curves upward by the same factor.

If the observed sub-PeV/PeV neutrinos originate from the sources of UHECRs,
the relevant region in the figure should overlap with the gray band.
 {
This implies that for Galactic-extragalactic transition at the second knee,
the pion production efficiency for escaping CRs must be $\sim 0.1$,
whereas it the transition is at the ankle, the efficiency must be even higher, i.e., $\gtrsim 0.5$.
}

 {
In plotting the various regions in Fig.~1,
we have assumed only representative values for each type of source,
without indicating the entire parameter space covered by that source class.
For all sources, we take a common range of values $\eta_p =0.01-1$
for the fraction of available energy that is channeled into escaping CR protons.
}

 {
Jets of active galactic nuclei (AGN) have long been considered one of the most promising candidates
for the sources of UHECRs as well as neutrinos \cite{Biermann87, Mannheim92}.
Here we consider only powerful objects with kinetic power $\sim 10^{45}\rm erg\,s^{-1}$
and source density $\sim 10^{-5}\rm Mpc^{-3}$ \cite{Berezinsky06,Ghisellini10},
which gives $\dot{W}_{0,\rm AGN}\sim \eta_p 10^{47.5}\rm erg\,Mpc^{-3}yr^{-1}$.
The pion production efficiency depends on the location of CR acceleration and neutrino production.
In the inner jet regions corresponding to the typical emission zones in blazars
within $\sim 10-100$ Schwarzschild radii of the central black hole,
the large photon density implies a high value, $0.1\lesssim f_\pi \leq 1$ \cite{Mannheim92, Mannheim01}
(note also \cite{Stecker91}).
In the outer jet regions such as the hot spots or radio lobes at kpc-Mpc scales
with much less ambient radiation, accordingly lower values are expected, $f_\pi\sim 10^{-3}-10^{-2}$
\cite{Biermann87, Anchordoqui08}.
These sites are respectively denoted ``AGN inner jets/cores" and ``AGN outer jets" in Fig.~1.
}

 {
Gamma-ray bursts (GRBs) have also been widely discussed as favorable sources of UHECRs \cite{Waxman95}.
Adopting an isotropic-equivalent kinetic energy per GRB of $10^{54}$ erg and a local GRB rate $\sim 1\,\rm Gpc^{-3}yr^{-1}$ \cite{Wanderman10}, we have $\dot{W}_{0,GRB}\sim \eta_p 10^{45}\rm erg\,Mpc^{-3}yr^{-1}$.
If CR acceleration occurs in the innermost regions of internal shocks with high photon density,
}
the pion production efficiency could be as high as $0.1\lesssim f_\pi \leq 1$ \cite{Waxman97},
 {
as indicated in Fig.~1 as ``GRB internal shocks".
Note, however, that the location of internal shocks can span a large range of radii
depending on the behavior of the central engine, and if it occurs in the outermost regions closer to the external shock,
much smaller values of $f_\pi$ are also possible.
}

 {
Clusters of galaxies, in particular the accretion shocks surrounding them,
have also been proposed as possible UHECR sources \cite{Norman95}.
Although it may be challenging to achieve maximum energies of $\sim 10^{20}$ eV,
acceleration up to $\gtrsim$ EeV may be quite feasible \cite{Inoue05}.
Furthermore, radio galaxies, i.e. AGN with jets, are sometimes found in the central regions of clusters,
which can also provide UHECRs inside clusters.
Such UHECRs can produce high-energy neutrinos via $pp$ collisions with the gas
constituting the intracluster medium (ICM) \citep{Murase08}.
If we consider massive clusters with $M \sim 10^{15} M_\odot$,
their space density is $\sim 10^{-6}\,\rm Mpc^{-3}$ and their expected accretion luminosity is $\sim 10^{46}\rm erg\,s^{-1}$,
so we arrive at $\dot{W}_{0,\rm IGS}\sim \eta_p10^{47.5}\rm erg\,Mpc^{-3}yr^{-1}$, comparable to that of AGN.
Assuming an average density of $10^{-4}\rm cm^{-3}$ for the ICM gas
and a residence time 1-10\,Gyr of high-energy protons inside the cluster,
we estimate a pion-production efficiency of $\sim 0.01-0.1$, outlined in Fig.~1 as ``Clusters of galaxies".
}

\begin{figure}
\resizebox{\hsize}{!}{\includegraphics{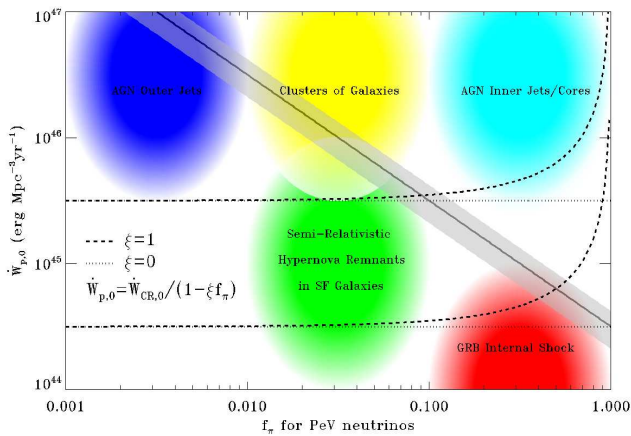}}
\caption{
 {The local proton energy production rate $\dot{W}_{p,0}$
versus $f_\pi$, the pion-production efficiency of escaping CRs that produce PeV neutrinos.
The black solid line represents the relation between $\dot{W}_{p,0}$ and $f_\pi$ required to reproduce
the observed neutrino flux, with the gray band corresponding to its 1-$\sigma$ confidence interval.
The upper and lower dashed curves represent the local energy production rate $\dot{W}_{\rm CR,0}$
of escaping CRs required to account for the observed UHECRs if the Galactic-extragalactic transition
occurs at the second knee and at the ankle, respectively, for the case $\xi=1$.
The dotted curves are corresponding ones for the case $\xi=0$. Here $\alpha=10$ is assumed for the normalization factor of the proton spectrum (note $\alpha={\rm ln} (E_{max}/E_{min})$
 for $s_p=2$). Larger/smaller values of alpha will shift all curves in the plot upwards/downwards by the same
 factor.
See text for discussion on the regions corresponding to different potential UHECR source candidates.}
\label{W-fpi}}
\end{figure}

Supernova remnants  (SNRs) have been widely discussed as promising accelerators of CR protons
(see \cite{Hillas05} for a review and references therein). However, standard treatments of shock acceleration in SNRs with ejecta {velocities $<10^9\rm cm\,s^{-1}$} reveal that
 {it is difficult to reach maximum energy $\gtrsim 40\left(\frac{1+z}{2}\right)\,$PeV,}
not to mention UHE protons with energy $\ge$EeV \cite{Lagage83}
(but see \cite{Ptuskin10} for discussions on acceleration during the very early stage of SNRs,
and \cite{Biermann88} on SNRs expanding into their progenitor winds).
However, a subset of very energetic supernovae called semi--relativistic hypernova (SR-hypernova),
has ejecta with much faster velocities, $\gtrsim 0.1c$, expanding into their progenitors' stellar winds \cite{Kulkarni98}.
Assuming a CR-amplified magnetic field with a strength close to equipartition,
SR-hypernovae satisfy the Hillas condition \cite{Hillas84} for acceleration 
of  $\gtrsim 10^{18}\,$eV  {protons} and have thus been proposed as sources of
UHECRs above the second knee \cite{Wang07}, or even up to the
highest  {CR energies when} considering the fastest part of the ejecta and
heavy nuclei acceleration \cite{Liu12}. SR-hypernovae are usually found associated with low-luminosity GRBs.
Although their event rate of $\sim 500\,\rm Gpc^{-3}yr^{-1}$
is lower than ordinary supernovae, the total kinetic energies released per event is larger, $\sim (3-5)\times 10^{52}\rm erg$ \cite{Kulkarni98}, providing a proton production rate $\dot{W}_{0,\rm HN}\sim \eta_p 10^{46}\rm erg\,Mpc^{-3}yr^{-1}$.
Below we estimate that the pion production efficiency for PeV neutrinos
 {due to {\it pp-}collisions between CRs escaping from SR-hypernova remnants and
the ambient interstellar medium (ISM) of their host galaxies is $\sim 0.1$,
although this is subject to uncertainties concerning the magnetic field and density of the host ISM.}
Thus, SR-hypernova remnants could be good candidates for the sources of the neutrinos detected by IceCube, as marked in Fig.~1.
Since Fig.~1 only describes a necessary condition for the link between IceCube neutrinos and UHECRs,
in the following sections we investigate in more detail whether a self-consistent model can be constructed
that ascribe the newly discovered sub-PeV/PeV neutrinos to SR-hypernovae remnants,
provided that they are also responsible for UHECRs above the second knee.
Since both Auger and HiRes indicate a rather light composition of UHECRs around the second knee \cite{composition},
we assume that
 {the source composition of CRs below $\sim 1$\,EeV is predominantly protons
and do not consider the effect of heavier nuclei in this paper.}

We point out that the marked regions in Fig.~1 for each source contain large uncertainties.
More precise values of $\dot{W}_{0,\rm CR}$ and $f_\pi$ depend on the details of the models.
Nonetheless, we can obtain a general idea about the plausibility of candidate sources.
As shown, if a certain type of source can only account for UHECRs above the ankle,
an extremely high pion-production efficiency (i.e. $f_{\pi}\simeq 1$) is needed to achieve sufficient PeV neutrino flux.
On the other hand, if the pion-production efficiency is too low (e.g.,$\lesssim 0.01$),
 {reproducing the observed neutrino flux requires a high proton production rate,
which in turn implies a low efficiency of CR escape from the sources to be consistent with the observed UHECR flux.}
We also note that Fig.~1 only gives constraints on some candidates from the viewpoint of the energy budget. These sources do not necessarily represent the common origins of these neutrinos and UHECRs even if they satisfy these energetics constraints.
 {Note also that some of these sources may already be constrained by other means.}
For instance, as indicated in \cite{Prev}, if the GRB internal shock model is responsible for the PeV neutrinos, IceCube should probably have already discovered a neutrino--GRB association both in time and space during its previous 40- and 59-string search \cite{IC59}.
 {Gamma-ray upper limits for some nearby, massive galaxy clusters
imply a low energy density of CRs at GeV-TeV energies in their ICM \cite{ZPP13},
which constrain their contributions to the diffuse neutrino background
at energies somewhat lower than those of the IceCube neutrinos.}
In simplest AGN models, $p\gamma$ collisions would lead to too many events at $\gtrsim$PeV energies, 
which is not favored by the current observation, unless extremely high magnetic field exists in the interaction region \cite{Winter13}.

In the SR-hypernova remnant model, the concomitantly produced isotropic gamma-ray flux may pose a potential problem.
Generally speaking,
 {if the diffuse neutrino flux is produced at the level of $10^{-8}\,\rm GeVcm^{-2}s^{-1}sr^{-1}$
via {\it pp-}collisions, the accompanying gamma-rays may overwhelm
the 0.1-100\,GeV diffuse isotropic gamma--ray background observed by Fermi/LAT \cite{Fermi10}
unless the source proton spectrum is sufficiently hard.}
As indicated in \cite{Murase13}, $s_p \gtrsim 2.2$ may already be in conflict with the gamma-ray background at low energies.
Although a hard spectrum of $s_p=2$ is employed in our calculation, we note that besides the proposed SR-hypernovae remnants,  {ordinary SNRs are expected to provide additional low-energy gamma-ray flux without contributing to sub-PeV/PeV neutrinos.}
Thus we must beware
that the total diffuse gamma-ray flux generated by SR-hypernova remnants and SNRs do not exceed the observed value.

\section{Neutrino emission from semi-relativistic hypernova remnants}
Accelerated protons from SR-hypernova remnants will interact with the ISM before escaping from
their host galaxies and produce 
neutrinos, gamma rays and electrons/positions. 
The energy loss time of CR protons in the ISM via  $pp$-collisions is
\begin{equation}
\begin{split}
\tau_{pp}(\varepsilon_p)&=[\, \kappa \sigma_{pp}(\varepsilon_p)nc\,]^{-1}\\
&=6\times 10^{7}{\rm yr}\,\left[\frac{\sigma_{pp}(\varepsilon_p={\rm 60\,PeV})}{100\,\rm mb}\right]^{-1}\left(\frac{n}{1\,\rm cm^{-3}}\right)^{-1}
\end{split}
\end{equation}
where $\kappa=0.17$ is the inelasticity, 
$\sigma_{pp}$ is the cross section, and $n$ is the
number density of  {ISM protons}. The $pp$-collision efficiency can be
estimated by $f_{\pi}={\rm min\,}(1,t_{\rm esc}/\tau_{pp})$ with
$t_{\rm esc}$  {as} the escape timescale.
Generally, there are two ways for  {CRs} to escape from a
galaxy. 
One, diffusive escape, is energy--dependent and the other,  advective escape via a galactic wind,
is energy--independent. 
The associated
escape timescales can be estimated by $t_{\rm diff}=h^2/4D$ and
$t_{\rm adv}=h/V_w$ respectively. 
Here $D=D_0(E/E_0)^{\delta}$ is
the diffusion coefficient where $D_0$ and $E_0$ are normalization
factors, and $\delta=0-1$ depending on  {the spectrum of interstellar magnetic turbulence}.
 $h$ is
usually taken as the scale height of the galaxy's gaseous disk
and $V_w$ is the velocity of the galactic wind in which the CRs are
advected. The  {diffuse gamma-ray emission from the Galactic plane} implies $f_\pi\sim 1\%$ for TeV protons \cite{Strong10},  {so} we may expect  {that} $f_\pi$ for $10$PeV protons  {is} $\ll 1\%$  in our Galaxy.
However, since the SR-hypernova rate  {should generally trace the cosmic star formation rate (SFR),
which is known to increase dramatically with $z$ from $z=0$ up to at least $z\sim\,$1--2 \cite{Hopkins06},
the properties of galaxies at $z\sim\,$1--2 (hereafter `high-redshift' galaxies) are likely to be more important
for determining the total diffuse neutrino flux.}
%
 {As our template systems, we consider high-redshift galaxies of two types,
normal star-forming galaxies (NSG) and starburst galaxies (SBG).}

High-redshift galaxies display different properties from nearby ones.
%
 {High-redshift NSGs generally do not reveal well-developed disk structure
and show more extended morphologies
with typical scale height $h\sim 1\,$kpc for massive systems} \cite{Daddi10,Law12}.
%
 {They also have much higher mass fractions of molecular gas}
\cite{Daddi10} with typical column density $\Sigma\sim
0.1\,$g\,cm$^{-2}$,  implying volumetric average ISM densities of $n\sim
\Sigma/2h\sim 10\,\rm cm^{-3}$.  
High-redshift  {SBGs} typically have scale height $h\sim 500\,$pc and
average gas density $n\sim 250\,\rm cm^{-3}$ \cite{Tacconi06}. 
As
for diffusion coefficients, recent studies on CR propagation
and anisotropy in our Galaxy suggest $D_0\sim 10^{28}\,\rm
cm^2s^{-1}$ at 3\,GeV and $\delta\simeq 0.3$ \cite{Trotta11}. 
 {Since little is}
known about the diffusion coefficient in high-$z$
galaxies,  we  {adopt} the same values of $D_0$ and $\delta$  {as inferred in our Galaxy
for high--redshift NSGs}. 
We assume a lower diffusion
coefficient $D_0\sim 10^{27}\,\rm cm^2s^{-1}$ for high-redshift
SBGs, because the magnetic fields in nearby SBGs such as M82 and
NGC253 are  {observed to be} $\sim 100$ times stronger than in our Galaxy and
the diffusion coefficient is expected to  {scale with the} CR's 
Larmor radius ($\propto \varepsilon_p/B$) \cite{Loeb06}. 
Regarding advective escape, the
velocity of the Galactic nuclear wind is $\sim 300\,\rm km s^{-1}$
\cite{Keeney06,
Crocker2012}, while optical and X-ray  {observations
show} the velocity of the outflow  {in} M82 are $\sim 500-600\, \rm
kms^{-1}$ \cite{McKeith95} and $1400-2200\,\rm kms^{-1}$
\cite{Strickland09} respectively. 
Since galactic winds are
probably driven by supernova explosions 
\cite{Chevalier85}  {whose rate is} higher in
high-redshift galaxies, we may expect  {their winds to be faster and}
take $V_w=500\,\rm
km\,s^{-1}$ and $1500\,\rm km\,s^{-1}$ as the reference values for
NSGs and SBGs respectively. 
Then we obtain
\begin{equation}\label{tdiffN}
t_{\rm diff}^{\rm N}=5\,\times 10^{4}{\rm yr}\, (\frac{h}{1\, \rm kpc})^2(\frac{D_0}{10^{28}\, \rm cm^2\,s^{-1}})^{-1}(\frac{\varepsilon_p}{60\,\rm PeV})^{-0.3}
\end{equation}
\begin{equation}\label{tadvN}
t_{\rm adv}^{\rm N}=2\,\times 10^{6}{\rm yr}\, (\frac{h}{1\, \rm kpc})(\frac{V_w}{500\, \rm km\,s^{-1}})^{-1}
\end{equation}
for  {NSGs}, and
\begin{equation}\label{tdiffB}
t_{\rm diff}^{\rm B}=10^{5}\,{\rm yr}\, (\frac{h}{0.5\, \rm kpc})^2(\frac{D_0}{10^{27}\, \rm cm^2\,s^{-1}})^{-1}(\frac{\varepsilon_p}{60\,\rm PeV})^{-0.3}
\end{equation}
\begin{equation}\label{tadvB}
t_{\rm adv}^{\rm B}=3\,\times 10^{5}{\rm yr}\, (\frac{h}{0.5\, \rm kpc})(\frac{V_w}{1500\, \rm km\,s^{-1}})^{-1}
\end{equation}
for  {SBGs}. 
The  {escape timescale can be approximated by}
$t_{\rm esc}={\rm min}\left(t_{\rm adv},\, t_{\rm
diff}\right)$,
and we may expect a break occurring in $t_{\rm esc}$ when $t_{\rm adv}=t_{\rm diff}$, i.e., 
$\varepsilon_{p,\rm b}^{\rm N}=300\,{\rm GeV}\,(\frac{h}{1\, \rm kpc})^{3.3}(\frac{V_w}{500\,\rm km\,s^{-1}})^{3.3}(\frac{D_0}{10^{28}\,\rm cm^2\,s^{-1}})^{-3.3}$
and
$\varepsilon_{p,\rm b}^{\rm B}=1.6\,{\rm PeV}\,(\frac{h}{1\, \rm kpc})^{3.3}(\frac{V_w}{1500\,\rm km\,s^{-1}})^{3.3}(\frac{D_0}{10^{27}\,\rm cm^2\,s^{-1}})^{-3.3}$ for  {NSGs} and SBGs respectively.
We then find that the $pp$-collision efficiencies for production of
1\,PeV neutrinos in  {NSGs} and SBGs are respectively
\begin{equation}
f_{\pi}^{\rm N}=t_{\rm
diff}^{\rm N}/\tau_{pp}^{\rm N}\simeq 0.01 ~{\rm and}~ f_{\pi}^{\rm B}=t_{\rm
diff}^{\rm B}/\tau_{pp}^{\rm B}\simeq
0.4
\end{equation}
The single-flavor neutrino flux at $1\,$PeV is
then $\varepsilon_\nu^2\Phi_\nu=\frac{1}{6}[f_{\rm SB}f_{\pi}^{\rm
B}+(1-f_{\rm SB})f_{\pi}^{\rm N}]\varepsilon_p^2\Phi_{\rm CR}\sim
10^{-8}\,\rm GeV\,cm^{-2}s^{-1}sr^{-1}$, which is comparable to
the observed neutrino flux. 
Here $f_{\rm SB}\sim 10\%-20\%$
\cite{Rodighiero11} is the fraction of the SFR contributed by
 {SBGs}. 
If we assume that SR-hypernovae account for CRs
above $\sim 5\times 10^{17}$eV, they should provide a CR flux of
$\varepsilon_p^2\Phi_{\rm CR}\simeq 7\times 10^{-7}\, \rm
GeV\,cm^{-2}s^{-1}sr^{-1}$ at this energy \cite{Hires04} and the required local
CR energy production rate $\dot{W}_0$  {is then} $ \sim
10^{45.5}\rm erg\,Mpc^{-3}yr^{-1}$.
 Assuming that each SR-hypernova
releases $E_{k,\rm HN}=5\times 10^{52}\,$erg  {of} kinetic energy \cite{Kulkarni98}, a
fraction  $\eta_p=$10\% of which goes into CRs, we find the
required local event rate is about $600\,\rm Gpc^{-3}yr^{-1}$,
consistent with the observed value \cite{Guetta07}.

The fluxes of secondary neutrinos and gamma rays produced by one
SR-hypernova $\phi_{\nu}$ and $\phi_{\gamma}$ (in unit of eV$^{-1}$)
are calculated  {with} the following analytical approximation \cite{Kelner06},
\begin{equation}
\phi_{i}(\varepsilon_i)\equiv \frac{dN_i}{d\varepsilon_i}\simeq \int_{\varepsilon_i}^{\infty}\frac{f_{\pi}}{\kappa}J_p(\varepsilon_p)F_i(\frac{\varepsilon_i}{\varepsilon_p},\varepsilon_p)\frac{d\varepsilon_p}{\varepsilon_p}
\end{equation}
where $i$ could be $\gamma$ or $\nu$. 
In the above equation, $F_i$
is the spectrum of  {the secondary $\gamma$ or $\nu$ in} a single collision. 
We assume that the accelerated
proton spectrum is $J_p=C_p
\varepsilon_p^{-2}{\rm exp}(-\varepsilon_p/\varepsilon_{p,\rm
max})$ where $C_p$ is a normalization coefficient  {fixed} by
$\int \varepsilon_p J_p d\varepsilon_p=\eta_pE_{k,\rm HN}$. 
Here we neglect the contribution of secondary electrons/positrons
and primary electrons via inverse Compton scattering and
Bremsstrahlung radiation, because these are only important at
$\lesssim 100\,$MeV \cite{Lacki12}.
To calculate the  {diffuse} flux of neutrinos and
gamma rays, we need to integrate the contribution from galaxies
throughout the whole universe, i.e.
\begin{equation}
\Phi_i(\varepsilon_i^{\rm ob})\equiv \frac{dN_i^{\rm ob}}{d\varepsilon_i^{\rm ob}}
=\frac{1}{4\pi}\int_0^{z_{\rm max}} \rho(z)\Gamma_{\rm HN}^{\rm SFR}\phi_i[(1+z)\varepsilon_i^{\rm ob}]\frac{cdz}{H(z)}
\end{equation}
where $\rho(z)=\rho_0S(z)$ represents the star-formation history with
$\rho_0$  being the  {local SFR} and $S(z)$
describing its evolution with redshift. 
The total SFR in the
local universe is found to be $\rho_0 \sim 0.01\,M_\odot\,\rm
yr^{-1}Mpc^{-3}$ and  {assumed to evolve} as \cite{Hopkins06} $S(z)\propto
(1+z)^{3.4}$ for $z<1$, $(1+z)^{0}$ for  $1 \leq z \leq 4$ and
$(1+z)^{-7}$ for  $z>4$. 
Here we assume the fraction of SFR from  {SBGs}
 {is} $f_{SB}=20\%$ at any cosmic epoch. 
The factor
$\Gamma_{\rm HN}^{\rm SFR}$ represents the ratio between the
SR-hypernova rate and SFR (in  {units} of $M_{\odot}^{-1}$). 
Its value is
normalized by requiring the local CR energy production rate of
 {SR-hypernovae to} match the observed CR flux above the second knee.
$H(z)=H_0\sqrt{\Omega_M(1+z)^3+\Omega_\Lambda}$ is the Hubble
 {parameter} and we adopt $H_0=71\, \rm km
s^{-1}Mpc^{-1}$, $\Omega_M=0.27$ and $\Omega_\Lambda=0.73$.
While neutrinos can reach the Earth without interaction, very 
high energy (VHE, $\gtrsim 100\,$GeV) gamma rays  {can} be absorbed 
by $e^{\pm}$ pair production on the intergalactic radiation field, 
initiating cascade processes and depositing  {energy} into $<100\,$GeV photons. 
As long as the 
cascade is well developed, the VHE gamma rays injected 
at $z$ will form a nearly universal  spectrum which only depends 
on the total energy injected and the injection redshift $z$ \cite{Coppi97}. 
We integrate over redshift to sum up the
contributions of cascades initiated at different $z$.

Panel (a) of Fig.~2 presents our calculated diffuse neutrino and gamma--ray fluxes.
The red dashed and dash-dotted lines
represent  {the} neutrino flux from  {NSGs} and SBGs respectively. 
At
low energies,  {energy--independent advective escape dominates over
energy-dependent diffusive escape}, so the spectrum of
neutrinos roughly follows the  {$s=-2$} accelerated proton spectrum. 
As  {the} energy
increases, the  neutrino spectrum breaks because  diffusive
escape becomes faster than advective escape.  
Because
$t_{\rm diff}\propto \varepsilon^{-0.3}$, the  {spectral index
above the break increases} by about 0.3.
But the
increase of the $pp$ cross section at higher energies
\cite{Kelner06} compensates  {this} somewhat, 
making the final spectral slope close to -2.2.
Note that in this case the UHECRs are mostly produced by
SR-hypernovae in NSGs while the PeV neutrinos mainly arise from
SR-hypernovae in SBGs. 
This is because most SR-hypernovae occur in
NSGs while the $pp$-collision efficiency is much higher in SBGs.

Given the uncertainties in $D$ at high redshift,
we also consider an alternative case in which $D_0$ in high-redshift NSGs is 
10 times smaller than in our Galaxy. 
There
is observational evidence for stronger magnetic fields in
such galaxies \cite{Bernet13}, 
so a smaller diffusion coefficient is plausible. 
With $D_0=10^{27}\rm
cm^{2}s^{-1}$ and assuming $f_{SB}=10\%$, we find that $f_{\pi}^{N}\simeq 0.1$ for production
of PeV neutrinos,  {in which case} both PeV neutrinos and UHECRs are produced
predominantly by hypernovae in  {NSGs}, as shown in panel (b) of Fig.~2.

If the observed sub-PeV/PeV neutrinos
 {originate from the sources of UHECRs},
their spectrum should extend to $\gtrsim 10\,$PeV without an abrupt cutoff.
In our model, the spectrum becomes softer at $\lesssim 10\,$PeV,
since the energy of the corresponding parent proton is $\lesssim 0.6(\frac{1+z}{2})\,$EeV,
approaching our assumed maximum energy  {of} $1\,$EeV.
 {This softening would not occur if $E_{p,\rm max}$ can be higher.}
Unless the propagation mode of CRs changes from diffusive to rectilinear above $\sim$EeV
and leads to a lower pion-production efficiency,
our model predicts a flux of a few times $10^{-9}\, \rm GeV\,cm^{-2}s^{-1}sr^{-1}$ around 10\,PeV,
as long as we assume the observed neutrinos and CRs above the second knee share a common origin.
This flux is consistent with
the present non-detection of neutrinos above several PeV, but is likely to be detectable in the future. 
Given that the all--flavor exposure of IceCube is $\sim 10^{15}\,\rm cm^{2}\,sr\,s$ at $10\,$PeV \cite{Anchordoqui13} for 662\,days, we may expect that such a flux of $\gtrsim 10\,$PeV neutrinos would be detected in $\lesssim 5-10$\,yrs operation.

\begin{figure}
\resizebox{\hsize}{!}{\includegraphics{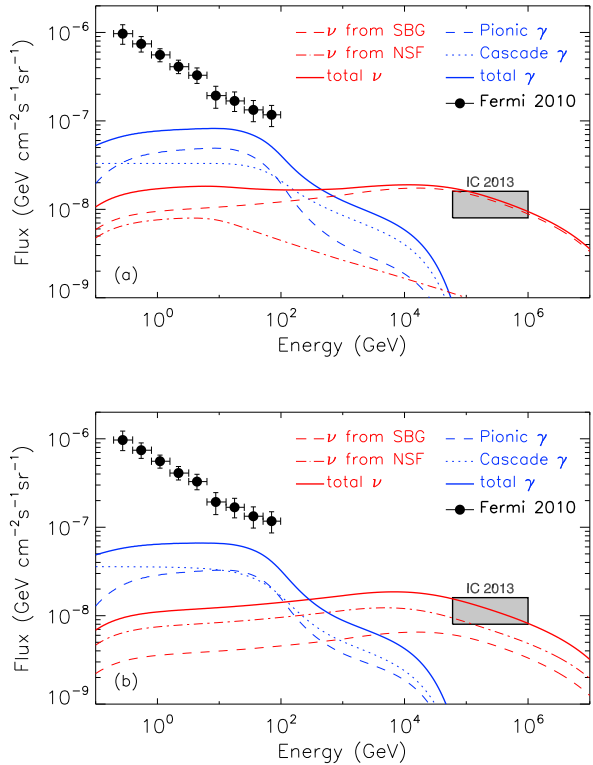}}
\caption{Spectra
of $\nu_\mu$ and gamma rays produced by SR-hypernova remnants in
star-forming galaxies. 
Upper panel: the red dashed line and dash--dotted line
represent the one--flavor neutrino flux from starburst galaxies
and normal star-forming galaxies respectively, and the red solid
line is  {their sum}. 
Neutrino oscillations imply that $\nu_\mu:\nu_e:\nu_\tau=1:1:1$ at the detector. 
The blue dashed and dotted lines represent the gamma ray fluxes from pion decay (accounting for intergalactic absorption) and the cascaded gamma ray flux, respectively, while the blue solid line is the sum of the two components. 
Data points are taken from \cite{Fermi10}.  
The shaded rectangle shows the  {IceCube flux \cite{ICSci13}}. 
Lower panel: same as the upper one but with $D_0=10^{27}\,\rm cm^2s^{-1}$ used for
normal star-forming galaxies and $f_{SB}=10\%$. See text for more discussion. 
\label{spectrum}}
\end{figure}

\section{Discussion} 
Including the cascade component, the total diffuse gamma ray flux
at $<100\,$GeV is  $\sim (7-8)\times 10^{-8}\,\rm
GeV\,cm^{-2}s^{-1}sr^{-1}$ in both cases, as shown with the solid blue lines in Fig.~2. 
Note that  putative additional losses due to 
absorption of VHE photons by
the radiation fields inside their host galaxies \cite{Inoue11} and by
synchrotron  {losses} of the $e^\pm$ pairs in the host galaxy
magnetic fields would lower the predicted cascade flux. 
The resulting flux is $\lesssim 50\%$ of the  {flux observed} by LAT. 
Also note that although normal  {SNRs} should not contribute to the $\gtrsim$100 TeV neutrino flux, they
can accelerate protons to PeV and produce $<100$\,TeV gamma rays,
contributing to the diffuse gamma-ray background. 
 {Compared to normal supernovae,
the local event rate of SR-hypernovae is $\sim 1\%$ while 
their explosion energy is dozens of times larger,} 
so the integral energy production rate of supernovae may be a few times larger than that of SR-hypernovae.
But the  {rate of SR-hypernovae relative to supernovae might be higher at high redshifts,
as SR-hypernovae may be engine-driven like long GRBs \cite{Paczynski98},
which} seem to occur preferentially in low-metallicity galaxies\cite{Stanek06}. 
This would suggest a relatively smaller contribution
 {of normal SNRs} at higher $z$. 
Nevertheless, as a rough estimate, we  may expect that
normal  {SNRs} could produce a gamma-ray flux
comparable to (or even less than) that of  SR-hypernova remnants, and in the former case the total gamma-ray
flux at 10\,--\,100\,GeV could reach the level of the observed one,
providing a possible explanation for the  {apparent} hardening in the spectrum
of the  {diffusive isotropic gamma-ray background at $>10\,$GeV}. On the other hand, we should also bear in mind that if the normal SNRs' energy budget turns out to be higher than that of SR-hypernovae even at high redshifts, the total generated diffusive gamma-ray flux would be a serious problem for this model.

As mentioned above, the spectral index of the high energy neutrino flux depends on the spectral indices of the injected protons and that of interstellar magnetic turbulence, i.e., $s_\nu\simeq s_p+\delta-0.1$ where $-0.1$ describes from the increase of the $pp$ cross section at high energy. Since the current measurement of the neutrino spectrum is far from accurate, if further observations show a different spectral shape, the values of $s_p$ or $\delta$ must be adjusted correspondingly. The most restrictive constraint on these two parameters comes from the concomitantly produced $<100\,$GeV gamma-ray flux in the pion-production process: this should not exceed the isotropic gamma-ray background observed by Fermi/LAT. Adopting either a larger $s_p$ or a larger $\delta$ would lead to a higher low-energy gamma-ray flux (see the Appendix for a detailed calculation). If future observations reveal a much softer neutrino spectrum, our model faces difficulties without invoking some untypical parameters or further refinements, e.g., introducing a break in the source spectrum. 

If SR-hypernovae are responsible not only for CRs above the second knee but also for those
at the highest energies, one may ask whether any of the neutrino events that have already been
observed by IceCube can be associated with individual sources within the GZK horizon of 100 Mpc \cite{GZK}.
According to our adopted evolution function $S(z)$ and assuming an isotropic sky distribution for such sources,
only about 0.3 out of the total of 28 events can be expected to come from within 100 Mpc.
In case such nearby sources happened to coincide with the direction of maximum effective area for IceCube,
then they may be responsible for about one of the 28 events.

The local SFR density is estimated to be $\sim 0.01\,M_\odot\,\rm Mpc^{-3}yr^{-1}$, and employing the
relation between SFR and infrared luminosity of a galaxy 
${\rm SFR}\,[M_\odot \,{\rm yr^{-1}}]=1.7\times 10^{-10} L_{\rm IR}
[L_\odot ]$ \cite{Magnelli11}, we find that a galaxy's CR
luminosity, accommodated by hypernovae, is $L_{\rm CR}\sim
10^{40}{\rm erg\,s^{-1}} (\dot{W}_0/10^{45.5}{\rm
erg\,Mpc^{-3}yr^{-1}})(L_{\rm IR}/10^{10}L_\odot)$. 
Given the infrared luminosity of our Galaxy is $\sim 10^{10}L_\odot$ and assuming a $pp-$collision efficiency of $10^{-3}$,  
we estimate the total Galactic neutrino luminosity at $100\,$TeV-1\,PeV is $\lesssim 10^{36}\rm erg\,s^{-1}$.
Note that our Galaxy might be too metal rich to host semi-relativistic hypernovae (or long GRBs) for  {the} last several billion years \cite{Stanek06}, so the real value could be smaller. Even if all these neutrinos are produced in the Galactic center and radiate isotropically,
it would result in $\lesssim 1$ event detection during 662\,days operation within a $8^\circ$ circular region around the Galactic Center \cite{Razzaque13} and would not cause a strong anisotropy that violates the observations \cite{ICSci13}. 

To summarize, we studied whether the newly-detected sub-PeV/PeV neutrinos can originate from the same sources as those responsible for CRs with energies above the second knee.
 {We discussed the conditions necessary for such a link between the observed PeV neutrinos and EeV CRs, and took SR-hypernova remnants in star-forming galaxies as an example of a self-consistent model that can provide the neutrino-UHECR link.
Comparing the predictions of different models,
the generated neutrino spectrum may vary somewhat from model to model,
and even within the same model depending on the uncertain parameters.
Thus, based on the spectral information alone,
SR-hypernova remnants can be neither confirmed nor refuted as the true sources of the observed neutrinos.
However, as long as the link between the observed neutrinos and EeV CRs exists,
we may generally expect detection of $\gtrsim 10\,$PeV neutrinos in the near future.
If such a link could be recognized, the detected neutrino flux and spectral shape should proffer information on the pion-production process at the sources. We shall then know that the real sources, whatever their identity, have a similar pion-production efficiency as that claimed for SR-hypernova remnants here, given that they also explain the UHECRs above the second knee. This provides us a chance to study the environment of the sources.
Future observations with greater statistics over the current neutrino energy range or detection at higher energies can give further constraints and help to uncover the true identity of the currently mysterious sources of EeV CRs.}

\acknowledgements
We thank the anonymous referees for useful comments, Luis Anchordoqui, Kohta Murase and Walter Winter for helpful discussions. This work is supported by the 973 program under grant 2014CB845800, the NSFC under grants 11273016 and 11033002, and the Excellent Youth Foundation of Jiangsu Province (BK2012011).

\appendix*
\section{Low-energy gamma-ray flux}
In the $pp$-collision process, neutrinos and neutral pions $\pi^0$ are produced as well as charged pions $\pi^\pm$, with branching ratios $\sim 1/3$. $\pi^0$ decays into two gamma rays, and each takes $\sim 10\%$ of the energy of the parent proton.
An approximate relation between the gamma-ray and neutrino flux from the same parent protons can be written as 
\begin{equation}\label{gamma-nu}
\varepsilon_\gamma^2\phi_\gamma(\varepsilon_\gamma= 2\varepsilon_\nu)=2\varepsilon_\nu^2\phi_\nu(\varepsilon_\nu)
\end{equation}
On the other hand, the relation between the neutrino flux at two different energies is 
\begin{equation}\label{nu-nu}
\varepsilon_\nu^2\phi_\nu(\varepsilon_{\nu,2})= [f_\pi(\varepsilon_{\nu,2})/f_\pi(\varepsilon_{\nu,1})](\varepsilon_{\nu,2}/\varepsilon_{\nu,1})^{2-s} \varepsilon_\nu^2\phi_\nu(\varepsilon_{\nu,1})
\end{equation}
Thus the gamma-ray flux at low energy, e.g., 10\,GeV, can be related to the neutrino flux at 1\,PeV as
\begin{equation}
\begin{split}
\varepsilon_{\gamma}^2\Phi_\gamma &|_{\tiny 10\,{\rm GeV}}=1.5(\frac{h}{0.5\,\rm kpc})^{-1}(\frac{V_w}{1500\,\rm kms^{-1}})^{-1}(\frac{D_0}{10^{27}\rm cm^2s^{-1}})\\
&\times(2\times 10^7)^{\delta -0.3}(2\times 10^5)^{s-2}\varepsilon_\nu^2\Phi_\nu |_{1\,{\rm PeV}}
\end{split}
\end{equation}
for $t_{\rm diff}<t_{\rm adv}<\tau_{pp}$, and
\begin{equation}
\begin{split}
\varepsilon_{\gamma}^2\Phi_\gamma &|_{10\,{\rm GeV}}=2.4(\frac{h}{0.5\,\rm kpc})^{-2}(\frac{n}{250\,\rm cm^{-3}})^{-1}(\frac{D_0}{10^{27}\rm cm^2s^{-1}})\\
&\times(2\times 10^7)^{\delta -0.3}(2\times 10^5)^{s-2}\varepsilon_\nu^2\Phi_\nu |_{1\,{\rm PeV}}
\end{split}
\end{equation}
for $t_{\rm diff}<\tau_{pp}<t_{\rm adv}$. The factor $2\times 10^7$ is the ratio between the energies of parent protons of 1\,PeV neutrinos and 10\,GeV photons, while the factor $2\times 10^5$ comes from substituting the value of the proton maximum energy $60\frac{1+z}{2}$\,PeV into the expression for the diffusion coefficient $D(\varepsilon_p)=D_0(\varepsilon_p/3\rm\,GeV)^\delta$. Here we have already taken $z=1$ for simplicity. Assuming the cascade of VHE gamma rays during propagation will double the GeV gamma-ray flux, we need $\varepsilon_{\gamma}^2\Phi_\gamma(10\,{\rm GeV})<10^{-7}\rm GeV\,cm^{-2}s^{-1}sr^{-1}$, i.e.,
\begin{equation}\label{sdelta1}
\begin{split}
&5.3\Delta s+7.3\Delta\delta \leq 0.52\\
&+{\rm lg}\,\left[(\frac{h}{0.5\,\rm kpc})(\frac{V_w}{1500\,\rm kms^{-1}})(\frac{D_0}{10^{27}\rm cm^2s^{-1}})^{-1}\right]
\end{split}
\end{equation}
for $t_{\rm diff}<t_{\rm adv}<\tau_{pp}$, and
\begin{equation}\label{sdelta2}
\begin{split}
&5.3\Delta s+7.3\Delta\delta \leq 0.32\\ 
&+{\rm lg}\,\left[(\frac{h}{0.5\,\rm kpc})^2(\frac{n}{250\,\rm cm^{-3}})(\frac{D_0}{10^{27}\rm cm^2s^{-1}})^{-1}\right]
\end{split}
\end{equation}
for $t_{\rm diff}<\tau_{pp}<t_{\rm adv}$ respectively, where $\Delta s=s-2$ and $\Delta \delta=\delta-0.3$. $t_{\rm diff}$ here is the diffusion escape time for the parent proton of a PeV neutrino. $t_{\rm diff}<t_{\rm adv}$ is usually true if typically expected values of $D_0$, $V_w$, $h$, $n$ are employed. Note that as we adjust the value of $s$ and $\delta$, these parameters also need to be changed in order to meet Eq.~(\ref{sdelta1}) and (\ref{sdelta2}). If future observations reveal a much softer neutrino spectrum, the changes in these parameters could be significant and some extreme values might be required. On the other hand, significant changes in these parameters could lead to $t_{\rm adv}<t_{\rm diff}$. In this case, the pion-production efficiency would hardly depend on the energy and directly implies $\Delta s\lesssim 0.13$. This would contradict the observed neutrino spectrum,
and hence our model would require further modifications, such as introducing a break in the source spectrum.

\end{document}